\documentclass[12pt]{article}
\usepackage[pctex32]{graphics}
\begin{document}
\vspace{1cm}
\begin{center}
~\\
{\bf  \Large Thermal Giant Graviton with Non-commutative Dipole Field}
\vspace{1cm}

                      Wung-Hong Huang\\
                       Department of Physics\\
                       National Cheng Kung University\\
                       Tainan, Taiwan\\

\end{center}
\vspace{1cm}
\begin{center}{\bf  \Large ABSTRACT } \end{center}
Using the type II near-extremal 3D-branes solution we  apply the T-duality and smeared twist to construct the supergravity backgrounds which dual to the 4D finite temperature non-commutative dipole field theories.  We first consider the zero-temperature system in which, depending on the property of dipole vectors it may be N=2, N=1 or N=0 theory.  We investigate the rotating D3-brane configurations moving on the spactimes and show that, for the cases of N=2 and N =1 the rotating D3-brane could be blowed up to the stable spherical configuration which is called as giant graviton and has a less energy than the point-like graviton.   The giant graviton configuration is stable only if its angular momentum was less than a critical value of $P_c$ which is an increasing function of the dipole strength. For the case of non-supersymmetric theory, however, the spherical configuration has a larger energy than the point-like graviton.  We also find that the dipole field always render the dual giant graviton to be more stable than the point-like graviton. The relation of dual giant graviton energy with its angular momentum, which in the AdS/CFT correspondence being the operator anomalous dimension  is obtained.  We furthermore show that the temperature does not change the property of the giant graviton, while it will render the dual giant graviton to be unstable.
\vspace{1cm}
\\
\\
\begin{flushleft}
E-mail:  whhwung@mail.ncku.edu.tw\\
\end{flushleft}

%%%%%%%%%%%%%%%%%%%%%%%

\newpage
\section{Introduction}
Giant graviton first investigated by  McGreevy, Susskind and Toumbas [1] is a rotating D3-brane in the $AdS_5 \times S^5$ spacetime, which is blowed up to the spherical BPS configuration and has the same energy and quantum number of the point-like graviton.  The configuration is stable only if its angular momentum was less than a critical value of $P_c$. This expanded brane wraps the spherical part of the $S^5$ spacetime and is stabilized against shrinking by the  of the Ramond-Ramond (RR) gauge field.  The authors in [2] had proved that the giant gravitons are BPS configurations which preserve the same supersymmetry as the point-like graviton.   As the giant graviton has exactly the same quantum numbers as the point-like graviton they can tunnel into each others.   It was also shown in [2] that there exist ``dual '' giant graviton consisting of spherical brane expanding into the AdS part of the spacetime, which, however, do not have an upper bound on their angular momentum due to the non-compact nature of the AdS spacetime.   While the giant graviton could tunnel into the trivial point-like graviton the investigations had shown that there is no direct tunneling between the giant graviton and its dual counterpart in AdS [3-5]. 

   The microscopical description of giant gravitons had also been investigated in [6,7].  The blowing up of gravitons into branes could  take place in backgrounds different from $AdS_5 \times S^5$ background, such as in the dilatonic background created by a stack of Dp-branes [8],  in the geometry created by a stack of non-threshold bound states of the type (D(p-2), Dp) [9],  in the  B-filed background [10], or in the Melvin field background [11].  The properties of the non-spherical giant was considered in [12].   The giant graviton solutions in Frolov's three parameter generalization of the Lunin-Maldacena background [13] had also been investigated in a recent paper [14,15].  In this paper we will investigate the giant gravitons on the supergravity backgrounds which dual to the 4D finite temperature non-commutative dipole field theories.

  In section II  we first construct the dual supergravity background of the finite temperature non-commutative dipole theory by considering the near-horizon geometry of near-extremal D-branes [16], after applying T-duality and  smeared as that described in [17,18].   Depending on the property of dipole vectors the theory may be N=2, N=1 or N=0 theory.  In section III we first investigate the zero temperature system and see that, for the cases of N=2 and N =1  a rotating D3-brane configurations moving on the spacetime could be blowed up to the stable spherical configuration.   We see that the giant graviton configuration is stable only if its angular momentum was less than a critical value of $P_c$ which is an increasing function of the dipole strength. For the case of non-supersymmetric theory, however, the spherical configuration has a larger energy than the point-like graviton.

 In section IV we first follow the Witten prescription [19]  to transform the supergravity background to the  global coordinate and then use the coordinate to study the dipole effect on the dual giant graviton.  It is found that the dipole field always renders the dual giant graviton more stable than the point-like graviton. We obtain the relation of dual giant graviton energy with its angular momentum, which in the AdS/CFT correspondence being the operator anomalous dimension.    In section V we turn to the finite temperature system and show that the temperature does not change the property of the giant graviton, while it will render the dual giant graviton to be unstable.  In last section we summary our results.

Note that the Giant graviton provides a very natural framework for the study of the gauge theory/gravity correspondence [20].  After studying the zero coupling limit of N = 4 super Yang-Mills theory with gauge group U(N) the candidate operators dual to giant gravitons had been proposed in [21-23].   Our investigations thus provides a correspondence which dual to a finite temperature non-commutative dipole field theories.   

%%%%%%%%%%%%%%%%%%%%%%%%%%%%
\section{Supergravity Solutions}
To find the explicit supergravity solution of D3-brane describing the finite temperature dipole theory we could start with the following type II supergravity 
solution describing \textsf{N} coincident near extremal D3-brane  [16] 
$$ds^2= f(r)^{-1/2}\left[- h(r)dt^2+dx_1^2+dx_2^2+dx_3^2\right] + f(r)^{1/2}\left[- h(r)^{-1}dr^2+ r^2 d\Omega_5^2\right],$$
$$ f(r) = 1+ {\textsf{N}^{4}\over r^4},~~~~~ h(r) = 1 - {r_0^{4}\over r^4}, \eqno{(2.1)}$$
in which $dr$ and $d\Omega$ constitute $x_4, ..., x_9$ coordinates. The horizon is located at $r =r_0$ and extremality is achieved in the limit $r_0 \rightarrow 0$. A solution with $r_0 \ll \textsf{N}$ is called near extremal.  

Now, as described in [17,18], we first apply the T-duality transformation on the $x_3$ axis. Then add a twist to the directions $x_4,....x_9$ as we go around the circle of new $x_3$ axis. This means that we replace
$$ dx_a \rightarrow  dx_a -\sum_{b=4}^9 \Omega_{ab}x_b dx_3,~~~~~ a= 4,...9,\eqno{(2.2)}$$
in which $\Omega$ is an element of the Lie algebra $SO(6)$. After the smeared twist along $x_4, ..., x_9$ we finally  apply the T-duality on the $x_3$ axis.   The supergravity solution becomes

$$ds^2 = f^{-\frac{1}{2}} \left( -h(r) dt^2 + dx^2_1 +dx^2_{2} 
+ \frac{ d x^2_3}{ 1 + r^2 n^T M^T M n} \right) \hspace{2cm}$$
$$+  f^{\frac{1}{2}} \left( h(r)^{-1} dr^2 + r^2 dn^T dn - \frac{r^4 (n^T M^T dn)^2}{1 + r^2 n^T M^T M n} \right), ,\eqno{(2.3)}$$
$$ e^{2\phi} =  \frac{1}{1 + r^2 n^T M^T M n }, ~~ \sum_{a=4}^9 B_{pa} dx_a = - \frac{r^2dn^TM n}{1  + 
r^2 n^T M^T M n},\eqno{(2.4)}$$
where $n$ is unit vector defined by $x_a=r n$ with $|n|^2=1$ and $\Omega \equiv e^{2\pi i M}$. The most general form of matrix $M$ can be cast to the following form [17]
$$ M=\pmatrix{0&\alpha_1+\alpha_2&0&0&0&0 \cr -\alpha_1-\alpha_2 
&0&0&0&0&0\cr 0&0&0&\alpha_1+\alpha_3&0&0 \cr 0&0& -\alpha_1
-\alpha_3 &0&0&0\cr 0&0&0&0&0&\alpha_2+\alpha_3 \cr 0&0&0&0& 
-\alpha_2-\alpha_3 &0}.\eqno{(2.5)}$$
This form of matrix $M$ breaks all supersymmetries.  On the other hand for $\alpha_3=0$ we left with 4 supercharges. For $\alpha_1=\alpha_2=0$ we find a configuration with 8 supercharges [17]. 
\\

 After the evaluations  the following supergravity solutions are found in the large \textsf{N} limit.

(1) $N=2$ theory:  We let  $\alpha_1=\alpha_2=0$ and $\alpha_3=B$ then 
$$ds_{10}^2 = U^2\left[- \left( 1-{U_T^4\over U^4}\right)dt^2+ dx^2+ dy^2+{ dz^2\over 1+B^2U^2\sin^2\theta}\right]+{1\over U^2} \left( 1-{U_T^4\over U^4}\right)^{-1}dU^2$$
$$+ d\theta^2 + \cos^2\theta d\phi^2 +\sin^2\theta\left(d\chi^2_1+\cos^2\chi_1 d\chi^2_2 + \sin^2\chi_1 d\chi^2_3\right)$$
$$ - {U^2B^2\sin^4\theta \left(\cos^2\chi_1 d\chi_2 + \sin^2\chi_1 d\chi_3\right)^2\over 1+U^2B^2\sin^2\theta}. \eqno{(2.6a)}$$
$$e^{2\Phi}= {1 \over  1+ U^2B^2\sin^2\theta},~~~
B_{z\chi_i}= - {U^2 B\sin^2\theta \left(\cos^2\chi_1 d\chi_2 + \sin^2\chi_1 d\chi_3\right) \over 1+U^2B^2\sin^2\theta}.\hspace{4cm}\eqno{(2.6b)}$$
Thus there is a nonzero B field with one leg along the brane worldvolume and others transverse to it.  The value $B$ in (2.6b) is proportional to the dipole length $\ell$ defined in the ``non-commutative dipole product" : $\Phi_a (x) * \Phi_a (x) = \Phi_a (x-\ell_b/2) ~\Phi_b (x +\ell_a/2) $ for the dipole field $\Phi(x)$ [17].  Note that the coordinate used  in (2.6a)  is like that in [1], which is slightly difference from that used in [24]. 
\\

(2) $N=1$ theory:  We let  $\alpha_1=\alpha_2= B$ and $\alpha_3=0$ then 
$$ds_{10}^2 = U^2\left[- \left( 1-{U_T^4\over U^4}\right)dt^2+ dx^2+ dy^2+{ dz^2\over 1+B^2U^2(3\cos^2\theta+1)}\right]+{1\over U^2} \left( 1-{U_T^4\over U^4}\right)^{-1}dU^2$$
$$+ d\theta^2 + \cos^2\theta d\phi^2 +\sin^2\theta\left(d\chi^2_1+\cos^2\chi_1 d\chi^2_2 + \sin^2\chi_1 d\chi^2_3\right) - {U^2B^2 \over 1+U^2B^2(3\cos^2\theta+1)}$$
$$\times \left[2\cos^2\theta d\phi + \sin^2\theta\left(\cos^2\chi_1 d\chi_2 + \sin^2\chi_1 d\chi_3\right)\right]^2. \eqno{(2.7a)}$$
$$e^{2\Phi}= {1 \over  1+U^2B^2(3\cos^2\theta+1)},\hspace{6.5cm}$$
$$(B_{z\phi},B_{z\chi_i})= -{ BU^2\left[2\cos^2\theta d\phi + \sin^2\theta\left(\cos^2\chi_1 d\chi_2 + \sin^2\chi_1 d\chi_3\right)\right]\over 1+U^2B^2(3\cos^2\theta+1)}.\eqno{(2.7b)}$$
\\

(3) $N=0$ theory:  We let  $\alpha_1=\alpha_2=\alpha_3= B/2$ then 
$$ds_{10}^2 = U^2\left[- \left( 1-{U_T^4\over U^4}\right)dt^2+ dx^2+ dy^2+{ dz^2\over 1+B^2U^2}\right]+{1\over U^2} \left( 1-{U_T^4\over U^4}\right)^{-1}dU^2$$
$$+ d\theta^2 + \cos^2\theta d\phi^2 +\sin^2\theta\left(d\chi^2_1+\cos^2\chi_1 d\chi^2_2 + \sin^2\chi_1 d\chi^2_3\right) - {U^2B^2 \over 1+U^2B^2}$$
$$\times \left[\cos^2\theta d\phi + \sin^2\theta\left(\cos^2\chi_1 d\chi_2 + \sin^2\chi_1 d\chi_3\right)\right]^2. \eqno{(2.8a)}$$
$$e^{2\Phi}= {1 \over  1+U^2B^2},~~(B_{z\phi},B_{z\chi_i})= -{ BU^2\left[\cos^2\theta d\phi + \sin^2\theta\left(\cos^2\chi_1 d\chi_2 + \sin^2\chi_1 d\chi_3\right)\right]\over 1+U^2B^2}.\eqno{(2.8b)}$$

In next section we first use the above geometries to study the giant graviton configurations at zero temperature.  The problem of finite temperature is studied in  section IV.

%%%%%%%%%%%%%%%%%%%%%%%%
\section{Zero-Temperature Giant Graviton with Dipole Field}
The rotating giant graviton we will search is the D3-brane wrapping the spherical $\chi_i$ spacetime.   The world-volume coordinate $\sigma_\mu$ are identified with the space-time coordinates by
$$\sigma_0= t,~~\sigma_1 = \chi_1,~~\sigma_2 = \chi_2,~~\sigma_3 = \chi_3,\eqno{(3.1a)}$$
and 
$$\phi=\phi(t).\eqno{(3.1b)}$$
The giant graviton will be fixed on the spatial coordinates at  $x=y=z=0$ with a fixed value of $U=1$.   To proceed, we know that the D3-brane action may be written as
$$S= -\int d^{4}\sigma~e^{-\Phi} \sqrt{-(g_{ab}+B_{ab})} +  \int P[A^{(4)}],\eqno{(3.2)}$$
where $g_{ab}$ ($B_{ab}$) is the pull-back of the spacetime metric ($B$ field) to the world-volume, and $P[A^{(4)}]$ denotes the pull-back of the $4$-form potential.   As the $B$ has only the component $B_{z\chi_i}$ or $B_{z\phi}$ (in N=0 case) it does not contribute to the Born-Infeld part of the action, as the world-volume coordinate of D3-brane is described by (3.1).

\subsection{N=2  Giant Graviton with Dipole Field}
Let us first the zero-temperature case of $N=2$ case.   As the RR field strength $F_{\theta \phi\chi_1\chi_2\chi_3}$ is proportional to $\sqrt {g_{\theta \phi\chi_1\chi_2\chi_3}}$ we can from the metric (2.6a) find that the 4-form RR potential on the dipole-field deformed $S^5$ is 
$$A^{(4)}_{\phi\chi_1\chi_2\chi_3} \approx \left(\sin^4\theta + {B^2\over 3} \sin^6\theta \right) \sin\chi_1\cos\chi_1 ,\eqno{(3.3)}$$
in which we consider only small dipole field theory to obtain an analytic form.  
The associated Lagrangian of the classical rotating D3-brane under ansatz  (3.1) is 
$$L \approx - \sin^3\theta \sqrt{1 -cos^2\theta~\dot\phi^2} + \left(\sin^4\theta + {B^2\over 3} \sin^6\theta \right)~\dot\phi, \eqno{(3.4)}$$
in which we have integrated the coordinate $\chi_i$. After the calculations the momentum conjugates to $\phi$ becomes
$$P \approx {cos^2\theta sin^3\theta~\dot\phi\over\sqrt{1 -cos^2\theta~\dot\phi^2}} + \sin^4\theta + {B^2\over 3} \sin^6\theta ,\eqno{(3.5)}$$
and associated energy of the dipole-field deformed giant graviton is 
$$H \approx {1\over cos\theta}  \sqrt{\left(P - \sin^4\theta - {B^2\over 3} \sin^6\theta\right)^2+ cos^2\theta~\sin^6\theta}.\eqno{(3.6)}$$

  Before using (3.6) to analyze the properties of the deformed giant graviton it is useful to know that the radius of an undeformed giant graviton is equal to its angular momentum, i.e. $ R=\sqrt P$ [1] (The radius of the giant graviton, $R$, is the value $sin\theta$ in our notation.).  Therefore, increasing the angular momentum of the undeformed giant graviton will increasing its size. In this case the giant graviton has same energy as the point-like graviton.  However, once its angular momentum is larger than 1, i.e. $P >1$, the configuration will has higher energy than that of the point-like graviton and giant graviton becomes unstable.
\\

To proceed, let us make the following comments:\\
1. The ``small value of $B$"  means that it is compared to the radius of the undeformed $S^5$ radius $R_{S}$.  Note that, for a convenience, we have let the radius $R_{S}=1$. \\
2.  Due to the deformation in the background caused by the dipole field the 
space described by the coordinate $\chi_i$ is not a sphere, as could be read from the metric form (2.6a).  Thus the giant gravitons which wrap around $\chi_i$ are not of spherical shape.  \\
3. As the giant graviton is not a sphere we could not use a radius to describe its shape.   However, using the metric form (2.6a) we could find its volume ($V_G$) and, for a convenience, we will use the ``effective radius" ($R_g$) to describe the giant graviton in which  $(R_g)^3 \sim ~ V_G $. The properties are also shown in the rest of paper.
\\

We could now use the formula (3.6) to plot the energy of the deformed giant graviton as a function of its effective radius $R_g$ (which is equal to $ R \equiv \sin\theta $) with various angular momentum $P$ under a fixed dipole field $B=0.3$.   The results are shown in figure 1. 
\\
 
\hfil\scalebox{1}{\includegraphics{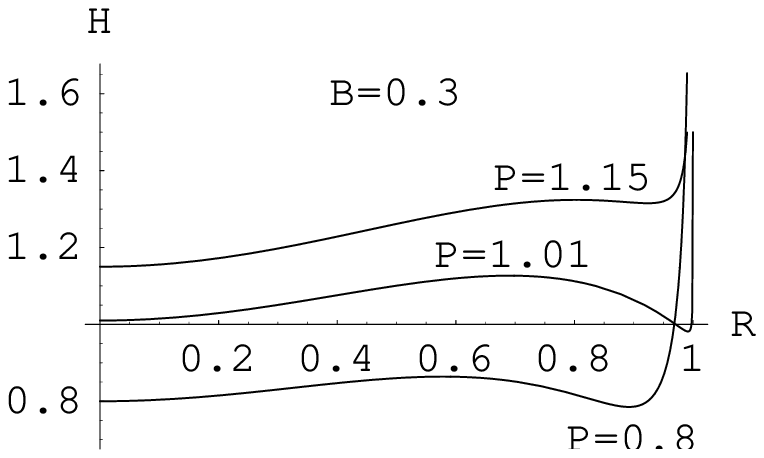}}\hfil\\
\\
{\it Figure 1:  Energy of the giant graviton with various angular momentum P=0.8, 1.01, or 1.05 as a function of its effective radius $R$ under a dipole field $B=0.3$. Giant graviton has maximum effective radius  $R=1$ at $P=P_c \approx 1.03$ and it becomes unstable as its angular momentum $P > P_c$.}
\\

 Note that the point-like graviton has radius $R=0$ and energy $H_0=P$ as could be easily read from (3.6).   In the case of small dipole field, we can also from (3.6) find that the effective radius and energy of the giant graviton are
$$R_g \approx  \sqrt P - {B^2\over 6}P^{5/2} .\eqno{(3.7)}$$
$$H(R_g) \approx P -  {B^2\over 3}P^3.\eqno{(3.8)}$$
Note that above relation implies that $H(R_g) - P < 0$ which, at first sight, is violating the ordinary BPS bound of undeformed theory [1,2].  In fact, for the dipole-field deformed theory the BPS inequality will be corrected.  Although the precise form, which may be derived form the dipole-field deformed supersymmetric algebra, remains to be investigated we would like to make an useful discussion in below.

   As we are considering the  case of  small dipole field the BPS inequality shall be slightly modified.  Therefore, in considering the configurations which are nearly the  giant graviton solution of (3.7), i.e. $R\approx R_g + \Delta R$, the energy of these states calculated from (3.6) will become
$$H(R)^2  \approx {1\over 1-R^2}  \left[\left(P - R^4 - {B^2\over 3}(R_g + \Delta R)^6 \right)^2+ (1-R^2) R^6 \right]$$
$$\approx {1\over 1-R^2}\left[\left(P - R^4 - {B^2\over 3}~ R_g^6 \right)^2+ (1-R^2) R^6 \right]$$
$$\approx  {1\over 1-R^2}  \left[\left(P - R^4 - {B^2\over 3} P^3 \right)^2+ (1-R^2) R^6 \right]$$
$$= \left(P- {B^2\over 3} P^3\right)^2 + {R^2\over (1-R^2)} \left(P - {B^2\over 3} P^3 - R^2  \right)^2 $$
$$= H(R_g)^2 + {R^2\over (1-R^2)} \left(P - {B^2\over 3} P^3 - R^2  \right)^2.\hspace{1.5cm} $$
This indicates that the giant graviton in the N=2 background may be a BPS configuration. The properties would also be shown in the rest of the paper (including the dual giant graviton) and, for short, we do not discuss them anymore.  However, to conclude that the giant graviton  is indeed a BPS state we need to  analyze the supersymmetry of  giant graviton [2].  The problem remains to be investigated. 

As the effective radius $R_g \equiv \sin\theta$ the condition of maximum value of $R_g=1$ then implies that
$$P_{c} \approx 1+ {B^2\over 3}.\eqno{(3.9)}$$
Thus the giant gravitons is a stable configuration if its angular momentum is less than $P_{c}$, which is an increasing function of the dipole strength.  For that case of $B = 0.3$, which is plotted in figure 1, we have $P_c \approx 1.03$.

 Thus, our analysis  have shown that the giant graviton may have lower energy than that of the point-like graviton.   This means that the dipole field could stabilize the giant graviton and suppress it from tunneling into the point-like graviton.

\subsection{N=1  Giant Graviton with Dipole Field}
The case of  $N=1$ could be analyzed in the same way. The associated Lagrangian of the classical rotating D3-brane under ansatz  (3.1) is 
$$L \approx - \sin^3\theta \sqrt{1 +4B^2 \cos^2\theta -\cos^2\theta~\dot\phi^2} + \left((1+ 4B^2) \sin^4\theta - 2B^2 \sin^6\theta \right)~\dot\phi, \eqno{(3.10)}$$
in which we have integrated the coordinate $\chi_i$. After the calculations 
the associated energy of the dipole-field deformed giant graviton is 
$$H \approx {\sqrt{1 +4B^2 \cos^2\theta}\over \cos\theta}  \sqrt{\left(P - (1+ 4B^2) \sin^4\theta + 2B^2 \sin^6\theta\right)^2+ \cos^2\theta~\sin^6\theta}.\eqno{(3.11)}$$

We could now use the above formula to plot the energy of the deformed giant graviton as a function of $\sin\theta$ with various angular momentum $P$ under a fixed dipole field.   The results are like those shown in figure 1 and giant graviton may have lower energy than that of the point-like graviton. 

 Note that the point-like graviton has radius $R=0$ and energy $H_0\approx P+2B^2P $ as could be easily read from (3.11).   In the case of small dipole field, we can also from (3.11) find that the effective radius and energy of the giant graviton are
$$R_g  \approx R \left(1+4B^2(1-R^2)\right)^{1/6},~~  R\equiv \sin\theta\approx  \sqrt P + B^2\left(5 P^{1/2} -10 P^{3/2}+ 4P^{5/2} \right).\eqno{(3.12)}$$
$$H(R_g) \approx P + 2B^2 (1-3P+P^2)P.\hspace{8.5cm}\eqno{(3.13)}$$
As the $R \equiv \sin\theta$ the condition of maximum value of $R=1$ then implies that
$$P_{c} \approx 1+ 2B^2.\eqno{(3.14)}$$
Thus the giant graviton with angular momentum $P< P_c$  will have lower energy than that of the point-like graviton and the dipole field could stabilize the giant graviton and suppress it from tunneling into the point-like graviton.

\subsection{N=0  Giant Graviton with Dipole Field}
The case of  $N=0$ could be analyzed in the same way. The associated Lagrangian of the classical rotating D3-brane under ansatz  (3.1) is 
$$L \approx - \sin^3\theta \sqrt{1 +B^2 \cos^2\theta -\cos^2\theta~\dot\phi^2} + (1- B^2) \sin^4\theta ~\dot\phi, \eqno{(3.15)}$$
in which we have integrated the coordinate $\chi_i$. After the calculations 
the associated energy of the dipole-field deformed giant graviton is 
$$H \approx {\sqrt{1 +B^2 \cos^2\theta}\over \cos\theta}  \sqrt{\left(P - (1- B^2) \sin^4\theta \right)^2+ \cos^2\theta~\sin^6\theta}.\eqno{(3.16)}$$

We could now use the above formula to plot the energy of the deformed giant graviton as a function of  $\sin\theta$  with various angular momentum $P$ under a fixed dipole field.   The results show that giant graviton always have larger  energy than that of the point-like graviton. 

 Note that the point-like graviton has radius $R=0$ and energy $H_0\approx P+B^2P/2 $ as could be easily read from (3.16).   In the case of small dipole field, we can also from (3.16) find that the effective radius and energy of the giant graviton are
$$R_g  \approx R \left(1+B^2(1-R^2)\right)^{1/6},~~  R\equiv \sin\theta \approx  \sqrt P + {B^2\over 4}\left(-3 P^{3/2} + 5 P^{3/2} \right).\eqno{(3.17)}$$
$$H(R_g) \approx P + {B^2\over2} (1+P)P.\hspace{8.5cm}\eqno{(3.18)}$$
As $H(R_g) > H_0 $  the giant graviton, if it exists, will always have a larger  energy than that of the point-like graviton. 

In conclusion, in this section we have shown that, for the cases of N=2 and N =1 the rotating D3-brane could be blowed up to the stable spherical configuration and it has a less energy than the point-like graviton.   The giant graviton configuration is stable only if its angular momentum was less than a critical value of $P_c$ which is an increasing function of the dipole strength. For the case of non-supersymmetric theory, however, the spherical configuration has a larger energy than the point-like graviton. 

%%%%%%%%%%%%%%%%%%%%%

\section{Zero-Temperature Dual Giant Graviton with Dipole Field}
In this section we first transform the supergravity solutions found in section II to the  global coordinate and then use the new coordinate to find the dual giant graviton solution.   We will see that the dipole field always render the dual giant graviton to be more stable than the point-like graviton.

\subsection{Metric}
To consider the dual giant graviton on the AdS [2] we will change the relevant coordinate to be the global coordinate following the Witten prescription [19].  Let use first investigate the following line element 
$$ds^2 = -(1+ r^2)d\tau^2 + r^2 \left({d\alpha_1^2\over 1+ r^2 B^4 \sin^2\theta} +\sin \alpha_1^2 d\alpha_2^2 +\sin\alpha_1^2 \sin\alpha_2^2 d\alpha_3^2\right) + {dr^2\over 1+ r^2}+ \cos\theta^2 d\phi^2. \eqno{(4.1)} $$
Using the new coordinates
$$U = B r, ~~~~  \tau = B t, \eqno{(4.2)} $$
then equation  (4.1) becomes
$$ds^2 \approx - U^2 dt^2 + U^2 {1\over B^2}\left({d\alpha_1^2\over 1+ U^2 B^2 \sin^2\theta} +\sin \alpha_1^2 d\alpha_2^2 +\sin\alpha_1^2 \sin\alpha_2^2 d\alpha_3^2\right) + {dU^2\over  U^2}+\cos\theta^2 d\phi^2, \eqno{(4.3)} $$
in the case of  $B \ll 1$.  As that described in [24], in the limit $B\rightarrow 0$ the original $S^3$ described by the coordinates  $\alpha_i$  become flat which may be described by $x,y,z$, as the radius ${1\over B}$ becomes infinite.  Thus the above line element could be written as 
$$ds^2 = \- U^2 dt^2 + U^2\left({dz\over 1+ U^2 B^2 \sin^2\theta} +dx^2 +dy^2\right) + {dU^2\over  U^2}+\cos\theta^2 d\phi^2,~~~~~~for~~N=2. \eqno{(4.4)} $$
This is the part of line element of  (2.6a) in the case of  $U_T=0$, which is the case of $N=2$.  Therefore we could use (4.1) to study the  dual giant graviton on the AdS in the case of small dipole field $B$.

   Through a  similar consideration we could find the relevant line elements for  the cases of $N=2$, $N=1$ and $N=0$ at finite temperature.  The results are
$$ds^2 = -\left( 1+r^2-{r_T^4\over r^2}\right)d\tau^2 + r^2 \left({d\alpha_1^2\over 1+ r^2 B^4 \sin^2\theta} +\sin \alpha_1^2 d\alpha_2^2 +\sin\alpha_1^2 \sin\alpha_2^2 d\alpha_3^2\right)\hspace{2cm} $$
$$+ {dr^2\over 1+r^2-{r_T^4\over r^2}}+ \cos\theta^2 d\phi^2, \hspace{4cm}~~~~~~~~~~for~~N=2,\eqno{(4.5)} $$
$$ds^2 = -\left( 1+r^2-{r_T^4\over r^2}\right)d\tau^2 + r^2 \left({d\alpha_1^2\over 1+ r^2 B^4(3\cos^2\theta+1)} +\sin \alpha_1^2 d\alpha_2^2 +\sin\alpha_1^2 \sin\alpha_2^2 d\alpha_3^2\right) \hspace{1cm}$$
$$+ {dr^2\over 1+r^2-{r_T^4\over r^2}} + \left[\cos^2\theta - {4 r^2B^4 \cos^4\theta\over 1+4^2B^4(3\cos^2\theta+1)}\right]d\phi^2,~~~~~~for~~N=1, \eqno{(4.6)}$$
$$ds^2 = -\left( 1+r^2-{r_T^4\over r^2}\right)d\tau^2 + r^2 \left({d\alpha_1^2\over 1+ r^2 B^4} +\sin \alpha_1^2 d\alpha_2^2 +\sin\alpha_1^2 \sin\alpha_2^2 d\alpha_3^2\right) \hspace{4cm}$$
$$+ {dr^2\over 1+r^2-{r_T^4\over r^2}} + \left[\cos^2\theta - { r^2B^4 \cos^4\theta\over 1+4^2B^4}\right]d\phi^2,\hspace{2cm}~~~~~~for~~N=0, \eqno{(4.7)}$$
in the case of small value of dipole field $B$.
%%%%%%%%%%%%%%%%%%%%%%%
\subsection{Dual Giant Graviton Solutions}
We can now use the above global coordinate  to investigate the dual giant graviton at zero temperature. The thermal dual giant graviton will be investigated in the next section

{\bf i) N=2}:  Using the ansatz [2]
$$\sigma_0 = \tau,~~\sigma_1 = \alpha_1,~~\sigma_2 = \alpha_2,~~\sigma_3 = \alpha_3,\eqno{(4.8a)}$$
and 
$$\phi=\phi(t),~~~\theta = \theta_0 .\eqno{(4.8b)}$$
the associated energy of the $N=2$ dual giant graviton calculated from (4.5) at $r_T=0$ is 
$$H \approx \sqrt{\left(1+r^2\right) \left(\cos^2\theta_0 P^2 + r^6\right)}~ - r^4 - B^4 r^6 {\sin^2\theta_0}\cos^4\theta_0 ,\eqno{(4.9)}$$
in the case of small value of dipole field $B$.  The solution of point-like graviton of radius $r=0$ has and energy $H(0) = P \cos\theta_0 $. 
\\

To proceed, let us make the following comments:\\
1. The ``small value of $B$"  means that it is compared to the radius of the undeformed AdS radius $R_{AdS}$.  Note that, for a convenience, we have let the radius $R_{AdS}=1$. \\
2.  Due to the deformation in the background caused by the dipole field the 
space described by the coordinate $\alpha_i$ is not a sphere, as could be read from the metric form (4.5).  Thus the dual giant gravitons which wrap around $\alpha_i$ are not of spherical shape.  \\
3. As the dual giant graviton is not a sphere we could not use a radius to describe its shape.   However, using the metric form (4.5) we could find its volume ($V_G$) and, for a convenience, we will use the ``effective radius" ($r_g$) to describe the dual giant graviton in which  $(r_g)^3 \sim ~ V_G $. The properties are also shown in the rest of paper.
\\

 The dual giant graviton effective radius and their associated energy are 
$$ r_g =  {r\over \left(1+r^2 B^2 \sin^2\theta_0\right)^{1/6}},~~ r \approx  \sqrt P \cos\theta_0 +  B^4 P^{3\over 2} \sin^2\theta_0 \cos^{3\over 2}\theta_0. \hspace{1cm}\eqno{(4.10)}$$
$$H(r_g) \approx P \cos\theta_0 - {13 B^4 P^3\over 3} \sin^2\theta_0 \cos^3\theta_0.\hspace{5.5cm}\eqno{(4.11)}$$
As $H(r_g) < H(0)$  the dual giant graviton will always have a less  energy than that of the point-like graviton. 

{\bf ii) N=1}:  Using  the ansatz (4.8) with $\theta_0=0$  the associated energy of the $N=1$ dual giant graviton calculated from (4.6) at $r_T=0$ is 
$$H \approx \sqrt{\left(1+r^2\right) \left(P^2 + r^6 - 4 r^4 B^4\right)}~ - r^4 -  {4\over 3} B^4 r^6,\eqno{(4.12)}$$
in the case of small value of dipole field $B$.  The solution of point-like graviton of radius $r=0$ has an energy $H(0) = P$.  The dual giant graviton effective radius  and their associated energy are 
$$r_g = {r\over \left(1+4r^2 B^2 \right)^{1/6}},~~ r \approx  \sqrt P +  B^4 P^{3\over 2}(5P + 6). \hspace{2cm}\eqno{(4.13)}$$
$$H(r_g) \approx P - {10 \over 3} B^4 P^3.\hspace{7cm}\eqno{(4.14)}$$
As $H(r_g) < H(0)$ the dual giant graviton will always have a less  energy than that of the point-like graviton. 

{\bf  iii) N=0}:  Using  the ansatz (4.8) with $\theta_0=0$  the associated energy of the $N=0$ dual giant graviton  calculated from (4.7) at $r_T=0$ is 
$$H \approx \sqrt{\left(1+r^2\right) \left(P^2 + r^6 -  r^4 B^4\right)}~ - r^4 -  {1\over 3} B^4 r^6,\eqno{(4.15)}$$
in the case of small value of dipole field $B$.  The solution of point-like graviton of radius $r=0$ has an energy $H(0) = P$.  The dual giant graviton effective radius  and their associated energy are 
$$ r_g= {r\over \left(1+r^2 B^2 \right)^{1/6}},~~r \approx  \sqrt P + {1\over 4} B^4 P^{3\over 2}(5P + 6). \hspace{1cm}\eqno{(4.16)}$$
$$H(r_g) \approx P - {5 \over 6} B^4 P^3.\hspace{7cm}\eqno{(4.17)}$$
As $H(r_g) < H(0)$ the dual giant graviton will always have a less  energy than that of the point-like graviton. 

In conclusion, we have shown that the dipole field always render the dual giant graviton to be more stable than the point-like graviton.  Note that in the AdS/CFT correspondence, we identify the energy  $H$  in global coordinates as 
the operator dimension in the field theory and $P$ as the R-charge, so we have computed the dimensions of operators corresponding to the brane configuration [20-22].

%%%%%%%%%%%%%%%%%%%%%
\section{Thermal Giant Graviton and Thermal Dual Giant Graviton}
\subsection{Thermal Giant Graviton }
Let us first consider the rotating D3-brane wraps the spherical part of $S^5$ without the dipole.  The relevant metric read from (2.6a) in the case of $B=0$ is 
$$ds^2 = - U^2\left( 1-{U_T^4\over U^4}\right)dt^2
+ d\theta^2 + \cos^2\theta d\phi^2 +\sin^2\theta\left(d\chi^2_1+\cos^2\chi_1 d\chi^2_2 + \sin^2\chi_1 d\chi^2_3\right). \eqno{(5.1)}$$
Using the ansatz  (3.1) the associated energy of the thermal giant graviton is 
$$H = ~ {\sqrt{1 - U_T^4}\over \cos\theta}  \sqrt{\left(P - \sin^4\theta \right)^2+ \cos^2\theta~\sin^6\theta},\eqno{(5.2)}$$
in which the giant graviton is moving along  $U=1$. As the temperature effect ($T = U_T/\pi$) only shows in an overall factor it thus does not affect the property of giant graviton.

In a similar way the thermal giant gravitons with dipole-field deformation have the following energy

$$H \approx  ~ {\sqrt{1 - U_T^4}\over cos\theta}  \sqrt{\left(P - \sin^4\theta - {B^2\over 3} \sin^6\theta\right)^2+ cos^2\theta~\sin^6\theta},~~~~N=2.\hspace{3.1cm}\eqno{(5.3)}$$
$$H \approx \sqrt{1 - U_T^4} ~ {\sqrt{1 +4B^2 \cos^2\theta}\over \cos\theta}  \sqrt{\left(P - (1+ 4B^2) \sin^4\theta + 2B^2 \sin^6\theta\right)^2+ \cos^2\theta~\sin^6\theta},~~~~N=1.\eqno{(5.4)}$$
$$H \approx \sqrt{1 - U_T^4} ~ {\sqrt{1 +B^2 \cos^2\theta}\over \cos\theta}  \sqrt{\left(P - (1- B^2) \sin^4\theta \right)^2+ \cos^2\theta~\sin^6\theta},~~~~N=0.\hspace{1.cm}\eqno{(5.5)}$$
\\

Comparing above equations with (3.6), (3.11), and (3.16) we see that the temperature effect ($T = U_T/\pi$) only shows in an overall factor it thus does not affect the property of giant graviton.  We therefore conclude that the temperature does not change the property of the giant graviton with dipole field.

\subsection{Thermal Dual Giant Graviton}
We next  consider the rotating D3-brane wraps the spherical part of $AdS^5$ without the dipole.  The relevant metric read from (4.5) with $B=0$ is
$$ds^2 = -\left( 1+r^2-{r_T^4\over r^2}\right) dt^2 + r^2 \left(d\alpha_1^2+\sin\alpha_1^2 d\alpha_2^2 + \sin\alpha_1^2  \sin\alpha_2^2 d\alpha_3^2 \right)+ {dr^2\over  1+r^2-{r_T^4\over r^2}} + d\phi^2  \eqno{(5.6)}$$
Using the ansatz [2]
$$\sigma_0= t,~~\sigma_1 = \alpha_1,~~\sigma_2 = \alpha_2,~~\sigma_3 = \alpha_3,\eqno{(5.7a)}$$
and 
$$\phi=\phi(t).\eqno{(5.7b)}$$
the associated energy of the thermal  dual giant graviton is 
$$H =\sqrt{\left(1+r^2-{r_T^4\over r^2}\right) \left(P^2 + r^6\right)}~ - r^4,\eqno{(5.8)}$$

In the case of  $r_T=0$ the dual giant graviton has radius $r_g=\sqrt P$ with   energy $H(r_g) = P$ which has the same value of the point-like graviton of $r=0$ [2]. 

 For the case of  $r_T\ne 0$, as the minimum value of $r_0$ satisfies the relation $1+r_0^2-{r_T^4\over r_0^2}=0$ the associated energy becomes $H(r_0) = -r_0^4$ which has a negative value.  However, for the dual giant graviton the associated energy is a positive value. (The property could be seen in the case of small $r_T$, in which the giant graviton energy is slightly different from $H(R_g) = P$.)

For a convenience we use (5.8)  to plot the energy of the dual giant graviton as a function of its radius $r$ with the angular momentum $P=5$ for the cases of $r_T=0$ and $r_T=1$.   The results are shown in figure 2. 
\\
 
\hfil\scalebox{1}{\includegraphics{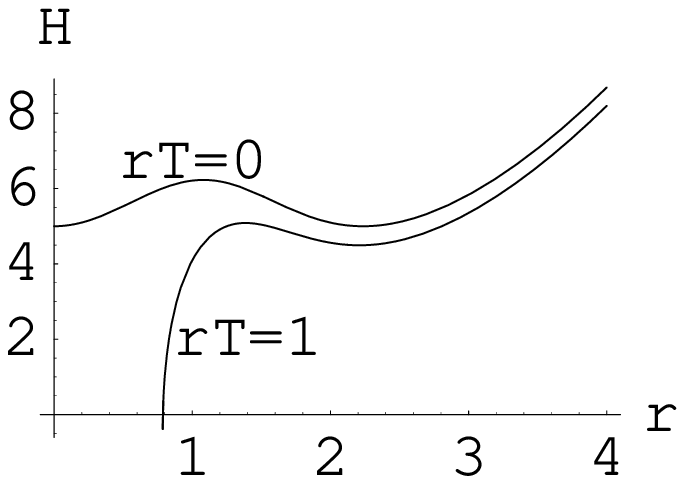}}\hfil\\
{\it Figure 2:  Energy of the dual giant graviton with angular momentum $P=5$ as a function of its radius $r$. At zero temperature ($r_T=0$) the dual giant graviton has the same energy as the point-like graviton.  However, at finite temperature ($r_T=1$) the configuration with a minimum value of $r_0~ (\approx 0.786)$ has a negative energy which is less then the dual giant graviton energy.}
\\

In a similar way  the thermal dual giant gravitons with small dipole-field deformation have the following energy
$$H \approx \sqrt{\left(1+r^2-{r_T^4\over r^2}\right) \left(\cos^2\theta_0 P^2 + r^6\right)}~ - r^4 - B^4 r^6 {\sin^2\theta_0}\cos^4\theta_0 ,~~~N=2.\eqno{(5.9)}$$
$$H \approx \sqrt{\left(1+r^2-{r_T^4\over r^2}\right) \left(P^2 + r^6 - 4 r^4 B^4\right)}~ - r^4 -  {4\over 3} B^4 r^6,~~~N=1.\hspace{1.5cm}\eqno{(5.10)}$$
$$H \approx \sqrt{\left(1+r^2-{r_T^4\over r^2}\right) \left(P^2 + r^6 -  r^4 B^4\right)}~ - r^4 -  {1\over 3} B^4 r^6,~~~~~N=0.\hspace{1cm}\eqno{(5.11)}$$
Through the same arguments we see that the energy of configuration with  a minimum radius $r_0$, which satisfies the relation $1+r_0^2-{r_T^4\over r_0^2}=0$, will  have a negative value while the energy of dual giant graviton  is a positive value.  We can therefore conclude that the temperature will render the dual giant graviton to be unstable.

%%%%%%%%%%%%%%%%%%%%%
\section{Conclusion}
Since Witten [19] had shown that the Ads-Schwarzschild spacetime could be used to study the dual finite temperature gauge theory many literatures has investigated the serval problems in the dual gravity side, including the finite
Wilson-Polyakov Loop [25,24].     In this paper, we study the giant graviton with a non-commutative dipole field deformation on the finite temperature system.  As the giant graviton provides a very natural framework for the study of the gauge theory/gravity correspondence [20] and the candidate operators dual to giant gravitons had been proposed in [21-23] our investigations thus provides a correspondence which dual to a finite temperature non-commutative dipole field theories.  

  In this paper, we use the type II near-extremal 3D-branes solution to construct the supergravity backgrounds by applying the T-duality and smeared twist, which dual to the 4D finite temperature non-commutative dipole field theories.  We have consider the zero-temperature system in which, depending on the property of dipole vectors it may be N=2, N=1 or N=0 theory.  We first show that, for the cases of N=2 and N =1 the rotating D3-brane could be blowed up to the stable spherical configuration which is called as giant graviton and has a less energy than the point-like graviton.   The giant graviton configuration is stable only if its angular momentum was less than a critical value of $P_c$ which is an increasing function of the dipole strength. For the case of non-supersymmetric theory, however, the spherical configuration has a larger energy than the point-like graviton.  

   We also transform the supergravity background to the  global coordinate following the Witten prescription [19] and find that the dipole field always render the dual giant graviton to be more stable than the point-like graviton. The relation of dual giant graviton energy with its angular momentum, which in the AdS/CFT correspondence being the operator anomalous dimension  is obtained.  We furthermore have considered the finite-temperature system and show that the temperature does not change the property of the giant graviton, while it will render the dual giant graviton to be unstable.  Finally, the thermal property of giant graviton on the other dipole field deformed background  [26] is interesting and remains to be  studied.
%%%%%%%%%%%%%%%%%%%%%%%

\newpage
%%%%%%%%%%%%%%%%%%%%%%%
{\bf  \Large REFERENCES}
\begin{enumerate}
\item  J. McGreevy, L. Susskind and N. Toumbas, ``Invasion of Giant Gravitons from Anti de Sitter Space'', JHEP  0006 (2000) 008 [hep-th/0003075].
\item  M. T. Grisaru, R. C. Myers and O. Tafjord, ``SUSY and Goliath'', JHEP  0008 (2000) 040 [hep-th/0008015]; A. Hashimoto, S. Hirano and N. Itzakhi, ``Large Branes in AdS and their Field Theory Dual," 0008 (2000) 051 (2000)  [hep-th/0008016].
\item S. Das, A. Jevicki and S. Mathur, ``Giant Gravitons, BPS Bounds and 
Noncommutativity," Phys. Rev. D63 (2001) 044001 [hep-th/0008088].
\item S. Das, A. Jevicki and S. Mathur, `` Vibration Modes of Giant Gravitons," Phys. Rev. D63 (2001) 024013 [hep-th/0009019],
\item  J. Lee, ``Tunneling between the giant gravitons in AdS5 x S5'', Phys.Rev. D64 (2001) 046012 [hep-th/0010191].
\item B. Janssen and Y. Lozano,``A Microscopical Description of Giant Gravitons,'' Nucl.Phys. B658 (2003) 281 [hep-th/0207199].
\item B. Janssen and Y. Lozano, ``Non-Abelian Giant Gravitons,'' Class. Quant. Grav. 20 (2003) S517 [hep-th/0212257]. 
\item  S. R. Das, S. P.. Trivedi, S. Vaidya,``Magnetic Moments of Branes and Giant Gravitons,''  JHEP 0010 (2000) 037 [hep-th/0008203].
\item  J. M. Camino,``Worldvolume Dynamics of Branes,'' [hep-th/0210249]. 
\item  J. M. Camino and A.V. Ramallo, ``Giant gravitons with NSNS B field," JHEP 0109 (2001) 012 [hep-th/0107142];  ``M-Theory Giant Gravitons with C field," Phys.Lett. B525 (2002) 337 [hep-th/0110096 ].
\item Wung-Hong Huang, ``Electric/Magnetic Field Deformed Giant Gravitons in Melvin Geometry'', Phys.Lett. B635 (2006) 141-147 [hep-th/0602019]; ``Spinning String and Giant Graviton in Electric/Magnetic Field Deformed $AdS_3 \times S^3 \times T^4$'', Phys.Rev. D73 (2006) 126010 [hep-th/0603198].
\item S. Prokushkin and M. M.  Sheikh-Jabbari, ``Squashed giants: Bound states of giant gravitons,'' JHEP 0407 (2004) 077  [hep-th/0406053]; A. Mikhailov, ``Giant gravitons from holomorphic," JHEP 0011(2000) 027 [hep-th/0010206]; ``Nonspherical giant gravitons and matrix theory," [hep-0208077].
\item O. Lunin and J.M. Maldacena, ``Deforming Field Theories with $U(1)\times U(1)$ global symmetry and their gravity duals,"  JHEP 0505 (2005) 033 [hep-th/0502086]; S.A. Frolov, ``Lax Pair for Strings in Lunin-Maldacena Background," JHEP 0505 (2005) 069 [hep-th/0503201].
\item R. de M. Koch, N. Ives, J. Smolic, M. Smolic, ``Unstable Giants," Phys. Rev. D73 (2006) 0604007 [hep-hep-th/0509007].
\item M. Pirrone, ``Giants On Deformed Backgrounds,"  JHEP 0612  (2006) 064 [hep-0609173]; S. D. Avramis, K. Sfetsos, D. Zoakos, ``Complex marginal deformations of D3-brane geometries, their Penrose limits and giant gravitons,"  arXiv:0704.2067 [hep-th]. 
\item G.T. Horowitz and A.~Strominger, ``Black strings and P-branes,'' Nucl. Phys. B  360 (1991) 197.
\item A. Bergman and O. J. Ganor,``Dipoles, Twists and Noncommutative Gauge Theory," JHEP 0010 (2000) 018 [hep-th/0008030]; A. Bergman, K. Dasgupta, O. J. Ganor, J. L. Karczmarek, and G. Rajesh,``Nonlocal Field Theories and their Gravity Duals," Phys.Rev. D65 (2002) 066005 [hep-th/0103090];   K. Dasgupta and M. M. Sheikh-Jabbari, ``Noncommutative Dipole Field Theories," JHEP 0202 (2002) 002 [hep-th/0112064]. 
\item M. Alishahiha and H. Yavartanoo,``Supergravity Description of the Large N Noncommutative Dipole Field Theories," JHEP 0204 (2002) 031 [hep-th/0202131].
\item E.~Witten, ``Anti-de Sitter space, thermal phase transition, and confinement in  gauge theories,'' Adv.\ Theor.\ Math.\ Phys.\   2 (1998) 505 [hep-th/9803131].
\item J.~M.~Maldacena, ``The large N limit of superconformal field theories and supergravity,'' Adv. Theor. Math.  Phys. 2 (1998) 231 [hep-th/9711200];
 S. S. Gubser, I. R. Klebanov and A.~M.~Polyakov, ``Gauge theory correlators from non-critical string theory,'' Phys. Lett. B428 (1998) 105 [hep-th/9802109]; E.~Witten, ``Anti-de Sitter space and holography,'' Adv. Theor. Math. Phys. 2 (1998) 253 [hep-th/9802150].
\item  V. Balasubramanian, M. Berkooz, A. Naqvi and M. Strassler", Giant Gravitons in Conformal Field Theory", JHEP 0204 (2002) 034 [hep-th/0107119];
S. Corley, A. Jevicki and S. Ramgoolam, ``Exact Correlators of Giant Gravitons from Dual N = 4 SYM Theory,'' Adv. Theor. Math. Phys. 5 (2002) 809 
2002  [hep-th/0111222].
\item  S. Corley and S. Ramgoolam, ``Finite Factorization equations and sum rules for BPS Correlators in N=4 SYM Theory,'', Nucl. Phys. B641 (2002) 131  [hep-th/0205221];  D. Berenstein, ``Shape and Holography: Studies of dual operators to giant gravitons," Nucl. Phys. B675 (2003) 179 [hep-th/0306090];  R. de M. Koch and R. Gwyn, ``Giant Graviton Correlators from Dual $SU(N)$ super Yang-Mills Theory," JHEP 0411 (2004) 081 [hep-th/0410236].
\item R. Roiban, ``On spin chains and field theories,'' JHEP 0409 (2004) 023 [hep-th/0312218]; D. Berenstein and  S. A. Cherkis,``Deformations of N=4 SYM and integrable spin chain models,'' Nucl.Phys. B702 (2004) 49 [hep-th/0405215
]; S.A. Frolov, R. Roiban, A.A. Tseytlin, ``Gauge-string duality for superconformal deformations of N=4 Super Yang-Mills theory,'' JHEP 0507 (2005) 045 [hep-th/0503192].
\item Wung-Hong Huang, ``Wilson-t'Hooft Loops in Finite-Temperature Non-commutative Dipole Field Theory from Dual Supergravity,''  Phys.Rev. D (2007), arXiv:0706.3663 [hep-th].
\item S.-J. Rey, S. Theisen and J.-T. Yee,   ``Wilson-Polyakov Loop at Finite Temperature in Large N Gauge Theory and Anti-de Sitter Supergravity,'' Nucl.Phys. B527 (1998) 171-186 [hep-th/9803135]; A. Brandhuber, N. Itzhaki, J. Sonnenschein and S. Yankielowicz,  ``Wilson Loops in the Large N Limit at Finite Temperature,'' Phys.Lett. B434 (1998) 36-40 [hep-th/9803137] ; A. Brandhuber, N. Itzhaki, J. Sonnenschein and S. Yankielowicz,  ``Wilson Loops, Confinement, and Phase Transitions in Large N Gauge Theories from Supergravity,'' JHEP 9806 (1998) 001 [hep-th/9803263].
\item U. Gursoy and C. Nunez   ,  ``Dipole Deformations of N=1 SYM and Supergravity backgrounds with U(1) X U(1) global symmetry ,'' Nucl.Phys. B725 (2005) 45-92  [hep-th/0505100 ]; N.P. Bobev, H. Dimov, R.C. Rashkov ,  ``Semiclassical Strings, Dipole Deformations of N=1 SYM and Decoupling of KK Modes,'' JHEP 0602 (2006) 064 [hep-th/0511216]; .
\end{enumerate}
\end{document}